%% file: main.tex
\documentclass[a4paper,USenglish,cleveref, autoref, thm-restate]{lipics-v2021}

\hideLIPIcs 

\usepackage[nocompress]{cite}

\title{Automated Complexity Analysis of Integer Programs via Triangular Weakly Non-Linear
  Loops (Short WST Version)} 

\titlerunning{Automated Complexity Analysis of Integer Programs via TWN-Loops} 
\author{Nils Lommen}{\and \url{https://verify.rwth-aachen.de/nlommen/}}{lommen@cs.rwth-aachen.de}{https://orcid.org/0000-0003-3187-9217}{}

\author{Eleanore Meyer}{\and \url{https://verify.rwth-aachen.de/emeyer/}}{eleanore.meyer@cs.rwth-aachen.de}{https://orcid.org/0000-0003-1038-4944}{}

\author{Jürgen Giesl}{LuFG Informatik 2, RWTH Aachen University, Aachen, Germany \and \url{https://verify.rwth-aachen.de/giesl/}}{giesl@informatik.rwth-aachen.de}{https://orcid.org/0000-0003-0283-8520}{}

\authorrunning{N.\ Lommen, E.\ Meyer, J.\ Giesl} 



\relatedversiondetails{See \cite{lommen2022AutomaticComplexityAnalysis}. Full version, including all proofs}{https://arxiv.org/abs/2205.08869}

\funding{funded by the Deutsche Forschungsgemeinschaft (DFG, German Research Foundation) - 235950644 (Project GI 274/6-2) and the DFG Research Training Group 2236 UnRAVeL}

\nolinenumbers 

\EventEditors{John Q.
	Open and Joan R.
	Access}
\EventNoEds{2}
\EventLongTitle{42nd Conference on Very Important Topics (CVIT 2016)}
\EventShortTitle{CVIT 2016}
\EventAcronym{CVIT}
\EventYear{2016}
\EventDate{December 24--27, 2016}
\EventLocation{Little Whinging, United Kingdom}
\EventLogo{}
\SeriesVolume{42}
\ArticleNo{23}

\usepackage[noline,noend,linesnumbered]{algorithm2e}
\usepackage{tikz}
\usetikzlibrary{shapes,calc,arrows,automata,decorations.pathmorphing}
\usepackage{wrapfig}

\newcommand{\emphit}[1]{\textit{#1}}

\newcommand{\braced}[1]{\lbrace #1 \rbrace}
\newcommand{\NN}{\mathbb{N}}
\newcommand{\ZZ}{\mathbb{Z}}
\newcommand{\QQ}{\mathbb{Q}}
\newcommand{\NNC}{\overline{\mathbb{N}}}

\newcommand{\update}{\eta}
\newcommand{\guard}{\tau}
\newcommand{\MRF}{\text{M}\Phi\text{RF}}
\newcommand{\MRFs}{\text{M}\Phi\text{RFs}}
\newcommand{\state}[0]{\sigma}
\newcommand{\State}{\Sigma}
\newcommand{\initial}{\sigma_0}
\newcommand{\Valuation}{\Sigma}

\newcommand{\location}{\ell}

\newcommand{\TSet}{\mathcal{T}}
\newcommand{\BoundSet}{\mathcal{B}}
\newcommand{\VSet}{\mathcal{V}}
\newcommand{\LSet}{\mathcal{L}}
\newcommand{\FormulaSet}{\mathcal{F}}

\newcommand{\Size}{{\mathcal{SB}}}

\newcommand{\true}{\texttt{true}}
\newcommand{\landau}{\mathcal{O}}
\newcommand{\tool}[1]{\textsf{#1}}
\newcommand{\KoAT}[0]{\tool{KoAT}}

\newcommand{\valuation}{\sigma}
\newcommand{\AtomSet}{\mathcal{A}}
\newcommand{\abs}[1]{|#1|}
\newcommand{\IntProgram}{(\VSet,\LSet,\location_0,\TSet)}
\renewcommand{\emptyset}{\varnothing}

\DeclareMathOperator{\rc}{rc}

\newcommand{\xvec}{\vec{x}}
\newcommand{\entry}{\mathcal{E}}
\newcommand{\pret}{r}

\newcommand{\loc}{{\mathcal{RB}_{\TSet'_>}}}
\newcommand{\glo}{{\mathcal{RB}}}
\newcommand{\IntLoop}{(\guard, \update)}
\newcommand{\indv}{d}
\newcommand{\PPEE}{\mathbb{PE}}
\newcommand{\cl}[1]{\texttt{cl}^n_{#1}}
\newcommand{\clExp}[2]{\texttt{cl}^{#2}_{#1}}

\newcommand{\timeboundterm}{
	\sup \braced{ n \in \NN \mid \exists \, (\location', \state').\; (\location_0, \state_0) \; (\rightarrow^*_{\TSet} \circ \rightarrow_t)^n \; (\location', \state') }
}

\newcommand{\sizeboundterm}{
	\braced{ |\state'(v)| \mid \exists\, \location' \in \LSet.
		\; (\location_0, \state_0) \; (\rightarrow^*_{\TSet} \circ \rightarrow_r) \; (\location', \state')}
}

\crefname{definition}{Def.}{Def.}
\crefname{example}{Ex.}{Ex.}
\crefname{appendix}{App.}{App.}
\crefname{ex}{Ex.}{Ex.}
\crefname{theorem}{Thm.}{Thm.}
\crefname{lemma}{Lemma}{Lemma}
\crefname{section}{Sect.}{Sect.}
\crefname{subsection}{Sect.}{Sect.}
\crefname{algorithm}{Alg.}{Alg.}
\crefname{corollary}{Cor.}{Cor.}
\crefname{figure}{Fig.}{Fig.}

\newcommand{\sth}{\operatorname{sth}}

\begin{document}
\maketitle

\begin{abstract}
	There exist several results on deciding termination and computing runtime bounds for \emph{triangular weakly non-linear loops} (twn-loops).
	We show how to use results on such subclasses of programs where complexity bounds are computable within incomplete approaches for complexity analysis of full integer programs.
	To this end, we present a novel modular approach which computes local runtime bounds for subprograms which can be transformed into twn-loops.
	These local runtime bounds are then lifted to global runtime bounds for the whole program.
	The power of our approach is shown by our implementation in the tool $\tool{KoAT}$ which analyzes complexity of programs where all other state-of-the-art tools fail.
\end{abstract}

\section{Introduction}
\label{sec-introduction}
Most approaches for automated complexity analysis of programs are based on incomplete techniques like ranking functions.
However, there also exist
subclasses of programs where termination is \emph{decidable}
and in \cite{hark2020PolynomialLoopsTermination} we presented the first subclass where
runtime bounds are 
\emph{computable}:
For \emph{triangular weakly non-linear loops}
(twn-loops),  there exist \emph{complete} techniques for analyzing termination and runtime complexity.
An example for a twn-loop is:
\begin{equation}
	\label{WhileExample}
	\textbf{while } (x_1^2 + x_3^5 < x_2 \, \wedge \, x_1 \neq 0) \textbf{ do }
	(x_1, x_2, x_3) \leftarrow (4\cdot x_1, \,	9\cdot x_2 - 8\cdot x_3^3, \, x_3) \quad
\end{equation}
Its guard is a propositional formula over (possibly \emph{non-linear}) polynomial inequations.
The update is \emph{triangular}, i.e., we can order the variables such that the update of any $x_i$ does not depend on the variables $x_1, \ldots, x_{i-1}$ with smaller indices.
So the restriction to triangular updates prohibits ``cyclic dependencies'' of variables (e.g., where the new values of $x_1$ and $x_2$ both depend on the old values of $x_1$ and $x_2$).
For example, a loop whose body consists of the assignment $(x_1, x_2) \leftarrow
(x_1 + x_2^2, x_2+1)$ is triangular, whereas a loop with the body
$(x_1, x_2) \leftarrow (x_1 + x_2^2, x_1+1)$
is not triangular.
From a practical point of view, the restriction to triangular loops seems quite natural.
For example, in \cite{STTT22}, $1511$ polynomial loops were extracted from the \emph{Termination Problems Data Base} \cite{tpdb}, the benchmark collection which is used at the annual \emph{Termination and Complexity Competition} \cite{termcomp}, and only $26$ of them were non-triangular.

Furthermore, the update is \emph{weakly non-linear}, i.e., no variable $x_i$ occurs non-linear in its own update.
So for example, a loop with the body $(x_1, x_2) \leftarrow 
(x_1 +x_2^2, x_2+1)$
is weakly non-linear, whereas a loop with the body $(x_1, x_2)
\leftarrow (x_1 \cdot x_2, x_2+1)$
is not.
With triangularity and weak non-linearity,
by handling one variable after the other, one can compute a \emph{closed form} which corresponds to applying the loop's update $n$ times.
 Using these closed forms, termination can be reduced to an existential formula over $\ZZ$ \cite{frohn2020TerminationPolynomialLoops} (whose validity is decidable for linear arithmetic and where SMT solvers often also prove (in)validity in the non-linear case).
 In this way, one can show that non-termination of twn-loops over $\ZZ$ is semi-decidable (and it is decidable over the real numbers).
 While termination of twn-loops over $\ZZ$ is not decidable, by using the closed forms, \cite{hark2020PolynomialLoopsTermination}
 presented a ``\emph{complete}'' complexity analysis technique.
More precisely,
for every twn-loop over $\ZZ$, it infers a polynomial which is an upper bound on
the runtime for all those inputs where the loop terminates.
So for all (possibly
non-linear) terminating twn-loops over $\ZZ$, this technique \emph{always} computes
polynomial runtime bounds.
In contrast, existing tools based on incomplete techniques for complexity analysis often fail for programs with non-linear arithmetic.

In \cite{brockschmidt2016AnalyzingRuntimeSize,Festschrift} we presented such an incomplete
modular technique for complexity analysis which uses individual ranking functions for different subprograms.
In this paper, we introduce a novel approach to automatically infer runtime bounds for
programs possibly consisting of multiple loops by han\-dling some subprograms as twn-loops and by using ranking functions for others.
Thus, complete complexity analysis techniques for subclasses of programs with non-linear arithmetic are combined with incomplete techniques based on ranking functions.

\section{Integer Programs}
\label{Integer Programs}
Let $\VSet$ be a set of variables.
The set of \emph{atoms} $\AtomSet(\VSet)$ consists of all inequations $p_1 < p_2$ for polynomials $p_1,p_2\in\ZZ[\VSet]$.
$\FormulaSet(\VSet)$ is the set of all propositional \emph{formulas} built from atoms $\AtomSet(\VSet)$, $\land$, and $\lor$.
In addition to ``$<$'', we also use ``$\geq$'', ``$=$'', ``$\neq$'', etc., and negations ``$\neg$'', which can be simulated by formulas (e.g., $p_1 \geq p_2$ is equivalent to $p_2 < p_1 + 1$ for integers).

For integer programs, we use a formalism based on transitions, which also allows us to represent \textbf{while}-programs like \eqref{WhileExample} easily.
Formally, an integer program is a tuple $\IntProgram$ with a finite set of variables $\VSet$, a finite set of locations $\LSet$, a fixed initial location $\location_0 \in \LSet$, and a finite set of transitions $\TSet$.
A \emph{transition} is a 4-tuple $(\location,\guard,\update,\location')$ with a \emph{start location} $\location\in\LSet$, \emph{target location} $\location'\in\LSet\setminus\braced{\location_0}$, \emph{guard} $\guard\in\FormulaSet(\VSet)$, and \emph{update} $\update: \VSet\rightarrow\ZZ[\VSet]$.
Our programs may have \emphit{non-deterministic branching}, i.e., the guards of several applicable transitions can be satisfied.
To simplify the presentation, we do not consider ``temporary'' variables (whose update is non-deterministic), but the approach can easily be extended accordingly (see \cite{lommen2022AutomaticComplexityAnalysis}).

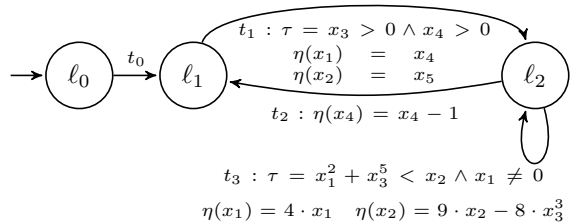
\begin{wrapfigure}[9]{r}{0.6\textwidth}
  \vspace*{-.3cm}
	\centering
	\begin{tikzpicture}[->,>=stealth',shorten >=1pt,auto,node distance=3.5cm,semithick,initial text=$ $]
		\node[state,initial] (q0) {$\location_0$}; \node[state] (q1) [right of=q0,xshift=-2cm]{$\location_1$}; \node[state,draw=none] (h0) [right of=q1, node distance=2cm]{}; \node[state] (q3) [right of=q1, node distance=4.5cm]{$\location_2$}; \draw (q0) edge node [text width=1.5cm,align=center] {{\scriptsize $t_0$}} (q1); \draw (q1) edge[bend left=75, looseness=.4] node [text width=4cm,align=center,below,yshift=-.1cm] {{\scriptsize $t_1:\guard = x_3 > 0 \wedge x_4 > 0$ \\
						$
							\begin{array}{rcl}
								\update(x_1) & = & x_4 \\
								\update(x_2) & = & x_5
							\end{array}
						$}} (q3); \draw (q3) edge[loop below, min distance=1cm] node [text width=5cm,align=center,xshift=-2cm,yshift=0.1cm]
			{{\scriptsize $t_3:\guard = x_1^2 + x_3^5 < x_2 \wedge x_1 \neq 0$ \\
						$\update(x_1) = 4\cdot x_1 \quad \update(x_2) = 9\cdot x_2 - 8\cdot x_3^3$}} (q3); \draw (q3) edge[bend left=10] node [text width=3cm,align=center]
			{{\scriptsize $t_2:\update(x_4) = x_4 - 1$}} (q1);
	\end{tikzpicture}
	\vspace*{-0.2cm}
	\caption{An Integer Program with a Nested Self-Loop}\label{fig:twnselfloop}
\end{wrapfigure}
\begin{example}
	Consider the program in \cref{fig:twnselfloop} with $\VSet = \{x_i \mid 1 \leq i \leq 5\}$, $\LSet = \{\location_0, \location_1, \location_2 \}$, and $\TSet = \braced{t_i \mid 0 \leq i \leq 3}$.
	We omitted trivial guards, i.e., $\guard = \true$, and identity updates of the form $\update(v) = v$.
	Here, $t_3$ corresponds to the \textbf{while}-program \eqref{WhileExample}.
\end{example}

A \emph{state} is a mapping $\valuation:\VSet\rightarrow\ZZ$, $\Valuation$ denotes the set of all states, and $\LSet\times\Valuation$ is the set of \emph{configurations}.
We also apply states to arithmetic expressions $p$ or formulas $\varphi$, where the number $\valuation(p)$ resp.\ the Boolean value $\valuation(\varphi)$ results from replacing each variable $v$ by $\valuation(v)$.
From now on, we fix a program $\IntProgram$.

\begin{definition}
	[Evaluation of Programs]\label{def:Evaluation}
	For configurations $(\location,\valuation)$, $(\location',\valuation')$ and $t = (\location_t,\guard,\update,\location_{t}')\in\TSet$, $(\location,\valuation)\rightarrow_t(\location',\valuation')$ is an \emph{evaluation}
	step if $\location = \location_t$, $\location' = \location_{t}'$, $\valuation(\guard) = \normalfont{\true}$, and $\valuation(\update(v)) = \valuation'(v)$ for all $v\in\VSet$.
	Let $\to_{\TSet} \; = \, \bigcup_{t \in \TSet} \to_t$, where we also write $\to$ instead of $\to_t$ or $\to_{\TSet}$.
	Let $(\location_0,\valuation_0)\rightarrow^k(\location_k,\valuation_k)$ abbreviate $(\location_0,\valuation_0)\rightarrow \ldots \rightarrow(\location_k,\valuation_k)$ and let $(\location,\valuation) \rightarrow^*(\location',\valuation')$ if $(\location,\valuation) \rightarrow^k(\location',\valuation')$ for some $k \geq 0$.
\end{definition}

So when denoting states $\valuation$ as tuples $(\valuation(x_1),\ldots,\valuation(x_5)) \in \ZZ^5$, for the program in \cref{fig:twnselfloop} we have $(\location_0,(7,5,1,1,3)) \to_{t_0}
	(\location_1,(7,5,1,1,3))\to_{t_1}
	(\location_2,(1,3,1,1,3)) \to_{t_3}^2 (\location_2,(16,163,1,1,3)) \ldots$
The \emph{runtime complexity} $\rc(\initial)$ of a program corre\-sponds to the length of the longest evaluation starting in the initial state $\initial$.
Formally, the runtime complexity is $\rc\!:\!\Valuation\!\rightarrow\!\NNC$ with $\NNC = \NN\cup\braced{\omega}$ and $\rc(\initial) = \sup\braced{k\!\in\!\NN\mid\exists (\location',\valuation').\, (\location_0,\initial)\!\rightarrow^k\!(\location',\valuation')}$.

\section{Computing Global Runtime Bounds for Integer Programs}
\label{Computing Global Runtime Bounds}

We now introduce our general approach for computing (upper) runtime bounds.
We use weakly monotonically increasing functions as bounds, since they can easily be ``composed'' (i.e., if $f$ and $g$ increase monotonically, then so does $f \circ g$).
The set of \emph{bounds} $\BoundSet$ is the smallest set with $\overline{\NN} \subseteq \BoundSet$, $\VSet\subseteq \BoundSet$, and $\{b_1+b_2, \, b_1 \cdot b_2, \, k^{b_1}\} \subseteq \BoundSet \text{ for all } k \in \NN$ and $b_1,b_2 \in \BoundSet$.
A bound constructed from $\NN$, $\VSet$, $+$, and $\cdot$ is \emph{polynomial}.
We measure the size of variables by their absolute values.
For any $\valuation \in \Valuation$, $\abs{\valuation}$ is the state with $\abs{\valuation}(v) = \abs{\valuation(v)}$ for all $v \in \VSet$.
So if $\initial$ denotes the initial state,
then $|\initial|$ maps every variable to its initial absolute value.
$\glo: \TSet \rightarrow \BoundSet$ is a \emph{global runtime bound}
if for each transition $t$ and initial state $\initial\in \Valuation$, $\glo(t)$ evaluated in the state $|\initial|$ over-approximates the number of evaluations of $t$ in any run starting in the configuration $(\location_0,\valuation_0)$.
So we have $|\initial|(\glo(t)) \; \geq \; \timeboundterm$ for all $t \in \TSet$ and all states $\initial \in \Valuation$ where $\rightarrow^*_{\TSet} \circ \rightarrow_t$ denotes the relation where arbitrary many evaluation steps are followed by a step with $t$.

For the program in \Cref{fig:twnselfloop}, we have $\glo(t_0) = 1$ (as $t_0$ is not on a cycle) and we will infer $\glo(t_i) = x_4$ for $i\in\braced{1,2}$ and $\glo(t_3) = x_4\cdot (2\cdot x_5 + 1)$ in \cref{ex:fullExample}.
By adding the bounds for all transitions, a global runtime bound $\glo$ yields an upper bound on the program's runtime complexity.
So for all $\initial\in\Valuation$ we have $|\initial|(\sum_{t\in\TSet}\glo(t)) \geq \rc(\initial).$

To infer global runtime bounds automatically, we first consider smaller subprograms $\TSet'\subseteq\TSet$ and compute \emph{local runtime bounds}.
A local runtime bound measures how often a transition $t\in\TSet'_>\subseteq\TSet'$ can occur in a run through $\TSet'$ that starts after an entry transition $\pret\in\entry_{\TSet'}$.
The \emph{entry transi\-tions} of $\TSet'$ are $ \entry_{\TSet'} = \braced{t \mid t\!=\!(\location,\guard,\update,\location')\!\in\!\TSet\setminus\TSet'\wedge \text{ there is a transition } (\location',\dots)\!\in\!\TSet'}$.
So in \cref{fig:twnselfloop}, we have $\entry_{\TSet\setminus\braced{t_0}} = \{ t_0 \}$ and $\entry_{\{t_3\}} = \{ t_1 \}$.
Thus, local runtime bounds do not consider how many $\TSet'$-runs take place in a global run and they do not consider the sizes of the variables before starting a $\TSet'$-run.
We lift these local bounds to global runtime bounds for the complete program afterwards.
Formally, $\loc\in\BoundSet$ is a \emph{local runtime bound} for $\TSet'_>$ w.r.t.\ $\TSet'$ if for all $t \in \TSet'_>$, all $\pret\in\entry_{\TSet'}$ with $r = (\location, \dots)$, and all $\state \in \State$, we have $|\state|(\loc) \geq\linebreak
	\sup \braced{ n \in \NN \mid \exists\, \initial, (\location', \state').
		\; (\location_0, \initial) \rightarrow_{\TSet}^* \circ \rightarrow_{\pret} \, (\location, \state) \; (\rightarrow_{\TSet'}^* \circ \rightarrow_t)^n \; (\location', \state') }$ for $\emptyset\neq\TSet'_>\subseteq \TSet'$.

\begin{example}
  In \cref{fig:twnselfloop}, by using the ranking function $x_4$, one can infer that
 $\mathcal{RB}_{\braced{t_1,t_2}} = x_4$ is
  a local runtime bound for $\TSet'_> = \braced{t_1,t_2}$ w.r.t.\ $\TSet' = \TSet'_>\cup
  \braced{t_3}$.
  For $\TSet'_> =\TSet' = \braced{t_4}$, our approach for twn-loops yields the
  local runtime bound  $\mathcal{RB}_{\braced{t_3}} = 2 \cdot x_2 + 1$, see \Cref{Local
    Runtime Bounds for TWN-Loops}.
\end{example}

If we have a local runtime bound $\loc$ w.r.t.\ $\TSet'$, then setting $\glo(t)$ to $\sum_{\pret \in \entry_{\TSet'}}
	\glo(\pret)\cdot (\loc \left[v/\Size(\pret,v) \mid v\!\in\!\VSet \right])$ for all $t\in\TSet'_>$ yields a global runtime bound.
Here, we over-approximate the number of local $\TSet'$-runs which are star\-ted by an entry transition $\pret \in \entry_{\TSet'}$ by an already computed global runtime bound $\glo(\pret)$.
Moreover, we instantiate each $v \in \VSet$ by a \emph{size bound} $\Size(\pret,v)$ which
is a bound on the size of $v$ before a local $\TSet'$-run is started.
To be precise, a size bound satisfies $|\initial|(\Size(r, v)) \geq \sup\sizeboundterm$ for all $(r, v) \in \TSet \times \VSet$ and all states $\initial \in\Valuation$.
Thus, our implementation alternates between runtime bound and size bound computations (see
\cite{brockschmidt2016AnalyzingRuntimeSize} for the computation of
size bounds).

\begin{example}
  \label{ex:fullExample}
  We now show how to obtain global runtime bounds from local runtime bounds in our
  example from \cref{fig:twnselfloop}. Here, we obtain the global runtime bound
        $\glo(t_1) = \glo(t_2) = \glo(t_0)\cdot (\glo_{\braced{t_1,t_2}}[v/\Size(t_0,v)
          \mid v \in \VSet	]) = x_4$ as $\Size(t_0,x_4) = x_4$, and
	 $\glo(t_3) = \glo(t_1)\cdot (\glo_{\braced{t_3}}[v/\Size(t_1,v) \mid v \in \VSet	]) = x_4\cdot(2\cdot x_5 + 1)$ as $\Size(t_1,x_2) = x_5$.
	Thus, $\rc(\initial) \in \landau(n^2)$ where $n$ is the largest initial absolute value of all variables.
	Our new technique allows us to use both local bounds resulting from twn-loops (for
        transition $t_3$ with non-linear arithmetic where tools based on ranking functions cannot infer a bound) and local bounds resulting from ranking functions (for $t_1$ and $t_2$).
\end{example}

To improve size and runtime bounds repeatedly, we treat the strongly connected components (SCCs)
of the program in topological order such that improved bounds for previous transitions are already available when handling the next SCC.
We first try to infer local runtime bounds by multiphase-linear ranking functions (see \cite{Festschrift} which also contains a heuristic for choosing $\TSet'_>$ and $\TSet'$ when using ranking functions).
If ranking functions do not yield finite local bounds for all transitions of the SCC, then we apply the twn-technique.
Afterwards, the global runtime bound is updated accordingly.
Note that the twn-approach is not only limited to self-loops but is also applicable for so-called simple cycles (see \cite{lommen2022AutomaticComplexityAnalysis}).
\section{Local Runtime Bounds for TWN-Loops}\label{Local Runtime Bounds for TWN-Loops}
In this section we briefly recapitulate how we infer runtime bounds for twn-loops, based
on \cite{frohn2020TerminationPolynomialLoops,hark2020PolynomialLoopsTermination}.
The tuple $(\guard,\update)$ is a twn-loop (over the variables $\vec{x} =
(x_1,\ldots,x_\indv)$) if $\guard\in\FormulaSet(\VSet)$ and $\update: \VSet \rightarrow
\ZZ[\VSet]$
for $\VSet = \{ x_1,\ldots,x_\indv \}$ such that for all $1\leq i \leq d$ we have
$\update(x_i) = c_i\cdot x_i + p_i$ for some $c_i\in\ZZ$ (where w.l.o.g.\ we can assume
$c_i \geq 0$) and $p_i\in\ZZ[x_{i+1},\dots,x_d]$.
Our algorithm starts with computing a closed form for the loop update, which describes the values of the variables after $n$ iterations of the loop.
Formally, a tuple of arithmetic expressions $\cl{\vec{x}} = (\cl{x_1}, \ldots,
\cl{x_\indv})$ over $\vec{x}$ and the distinguished variable $n$ is a \emph{closed form}
for the update $\update$ with start value $n_0 \geq 0$ if for all $1 \leq i \leq \indv$ and all $\sigma: \{x_1,\ldots,x_\indv,n\} \to \ZZ$ with $\sigma(n) \geq n_0$, we have $\sigma(\cl{x_i}) = \sigma(\update^n(x_i))$.
These closed forms can be represented as so-called \emph{poly-exponential expressions}.
The set of all poly-exponential expressions is defined as $\PPEE = \{ \sum_{j=1}^\ell p_j
\cdot n^{a_j} \cdot b_j^n \mathrel{\Big|} \ell, a_j\in \NN, \; p_j\in\QQ[\VSet], \;
b_j\in\NN_{\geq 1} \}$.

\begin{example}
	\label{ex:closed form}
	A closed form (with start value $n_0 = 0$) for the twn-loop in \eqref{WhileExample} is $\cl{x_1} = x_1 \cdot 4^n$, $\cl{x_2} = (x_2 - x_3^3) \cdot 9^n + x_3^3$, and $\cl{x_3} = x_3$.
\end{example}

Using the closed form, as in \cite{frohn2020TerminationPolynomialLoops} one can represent non-termina\-tion of a twn-loop $\IntLoop$ by the formula $\exists\, \xvec\in \ZZ^d, \; m\in \NN. \; \forall n \in \NN_{\geq m}. \; \guard[\vec{x}/\clExp{\vec{x}}{n}]$.
Here, $\guard[\vec{x}/\clExp{\vec{x}}{n}]$ means that each variable $x_i$ in $\guard$ is replaced by $\clExp{x_i}{n}$.
Hence, whenever $\forall n \in \NN_{\geq m}. \; \guard[\vec{x}/\clExp{\vec{x}}{n}]$ holds, then $\clExp{\vec{x}}{\max\{n_0,m\}}$ witnesses non-termination.
Thus, invalidity of the previous formula is equivalent to termination of the loop.
Poly-exponential expressions have the advantage that it is always clear which addend determines their asymptotic growth when increasing $n$.
So as in \cite{frohn2020TerminationPolynomialLoops}, the formula can be transformed into an existential formula and we use an SMT solver to prove its invalidity in order to prove termination of the loop.

As observed in \cite{hark2020PolynomialLoopsTermination}, since the closed forms for twn-loops are poly-exponential expressions that are weakly monotonic in $n$, every twn-loop $\IntLoop$ \emph{stabilizes} for each
input $\vec{e} \in \ZZ^d$.
So there is a number of loop iterations (a \emph{stabilization threshold} $\sth_{\IntLoop}(\vec{e})$), such that the truth value of the loop guard $\guard$ does not change anymore when performing further loop iterations.
Hence, the runtime of every terminating twn-loop is bounded by its stabilization threshold.
See \cite{hark2020PolynomialLoopsTermination} for the computation of bounds on the
stabilization thresholds.

\begin{example}
	The stabilization threshold $\sth_{\IntLoop}$ of the twn-loop \eqref{WhileExample}
        is bounded by $2\cdot x_2 + 1$.
	Thus, as the twn-loop \eqref{WhileExample} terminates, $2\cdot x_2 + 1$ is a bound on the runtime for the loop \eqref{WhileExample}.
	Due to the correspondence between $t_3$ and the loop \eqref{WhileExample},
        $2\cdot x_2 + 1$ is also a local runtime bound for $t_3$.
	\end{example}
\section{Conclusion and Evaluation}
\label{Evaluation}

We showed that results on subclasses of programs with computable complexity bounds like \cite{frohn2020TerminationPolynomialLoops,hark2020PolynomialLoopsTermination} are not only theoretically interesting, but they have an important practical value.
By integrating such complete techniques into incomplete approaches for general programs,
the power of automated complexity analysis is increased substantially, in particular because now one can also infer runtime bounds for programs containing non-linear arithmetic.
We evaluated this integration in our re-implementation of the tool \tool{KoAT} and compared the results to other state-of-the-art tools.
Let \tool{KoAT1} refer to the original tool from \cite{brockschmidt2016AnalyzingRuntimeSize} and let \tool{KoAT2} refer to our new re-implementation \cite{Festschrift}.
We tested the following configurations of \tool{KoAT2}, which differ in the techniques used for the computation of local runtime bounds:

\medskip
\hspace*{-.75cm}
\begin{minipage}{14cm}
	\begin{itemize}
		\item[$\bullet\!$] {\tool{KoAT2\,\!+\,\!$\MRF 5$} uses multiphase-linear ranking functions ($\MRFs$) of depth $\leq 5$}
		\item[$\bullet\!$] \tool{KoAT2\,\!+\,\!TWN} only uses the twn-technique
		\item[$\bullet\!$] \tool{KoAT2\,\!+\,\!TWN\,\!+\,\!$\MRF 5$} uses the twn-technique and $\MRFs$ of depth $\leq 5$
	\end{itemize}
\end{minipage}

\medskip

\begin{figure}[t]
	\makebox[\textwidth][c]{
		\begin{tabular}{l|c|c|c|c|c|c||c|c}
			                                                   & $\landau(1)$ & $\landau(n)$ & $\landau(n^2)$ & $\landau(n^{>2})$ & {\scriptsize $\landau(\mathit{EXP})$} & $< \infty$ & $\mathrm{AVG^+(s)}$ & $\mathrm{AVG(s)}$ \\
			\hline {\scriptsize \tool{KoAT2 + TWN + $\MRF 5$}} & 26           & 231          & 73             & 13                & 1                                     & 344        & 8.72                & 23.93             \\
			\hline \tool{KoAT2 + $\MRF 5$}                     & 24           & 226          & 68             & 10                & 0                                     & 328        & 8.23                & 21.63             \\
			\hline \tool{MaxCore}                              & 23           & 216          & 66             & 7                 & 0                                     & 312        & 2.02                & 5.31              \\
			\hline \tool{CoFloCo}                              & 22           & 196          & 66             & 5                 & 0                                     & 289        & 0.62                & 2.66              \\
			\hline \tool{KoAT1}                                & 25           & 169          & 74             & 12                & 6                                     & 286        & 1.77                & 2.77              \\
			\hline \tool{Loopus}                               & 17           & 170          & 49             & 5                 & 0                                     & 241        & 0.42                & 0.43              \\
			\hline \tool{KoAT2 + TWN}                          & 20           & 111          & 3              & 2                 & 0                                     & 136        & 2.54                & 26.59
		\end{tabular}
	}
	\caption{Evaluation on the Collection \tool{CINT} \vspace*{-.4cm}}
	\label{fig:CINT}
\end{figure}

\Cref{fig:CINT} presents our evaluation on the 504 \emph{Complexity C Integer Programs}
(\tool{CINT}) used in the annual \emph{Termination and Complexity Competition} \cite{termcomp}.
Here, all variables are interpreted as integers over $\ZZ$ (i.e., without overflows).
We compare \tool{KoAT} with the tools \tool{CoFloCo} \cite{flores-montoya2014ResourceAnalysisComplex}, \tool{MaxCore} \cite{albert2019ResourceAnalysisDriven} with \tool{CoFloCo} in the backend, and \tool{Loopus} \cite{sinn2017ComplexityResourceBound}.
In \Cref{fig:CINT}, the runtime bounds are compared asymptotically.
So for instance, there are $26 + 231 = 257$ programs in \tool{CINT} where \tool{KoAT2\,\!+\,\!TWN\,\!+\,\!$\MRF 5$} can show that $\rc(\valuation_0) \in \landau(n)$ holds for all initial states $\valuation_0$ where $\abs{\valuation_0(v)} \leq n$ for all $v \in \VSet$.
For $26$ of these programs, \tool{KoAT2\,\!+\,\!TWN\,\!+\,\!$\MRF 5$} can even show that $\rc(\valuation_0) \in \landau(1)$, i.e., their runtime complexity is constant.
Overall, this configuration succeeds on $344$ examples, i.e., ``$< \infty$'' is the number
of examples where a finite bound on the runtime complexity could be computed by the
respective tool within the time limit of 5 minutes per example.
``$\mathrm{AVG^+(s)}$'' is the average runtime of the tool on successful runs in seconds, i.e., where the tool inferred a finite time bound before reaching the timeout, whereas ``$\mathrm{AVG(s)}$'' is the average runtime of the tool on all runs including timeouts.

\KoAT's source code, a binary, and a Docker image are available at \url{https://koat.verify.rwth-aachen.de/twn}.
The website also has details on our experiments and \emph{web interfaces} to run \KoAT's configurations directly online.


\input{main.bbl}

\end{document}

%% file: main.bbl
\providecommand{\noopsort}[1]{}